\documentclass[aps,twocolumn,amsmath,amssymb,showpacs]{revtex4}

\usepackage{graphicx}
\usepackage{dcolumn}
\usepackage{bm}

\usepackage{epstopdf} 
\begin{document}

\title{Interaction Effects and Pseudogap in Two-Dimensional Lateral Tunnel Junctions}

\author{P. Jiang$^{*}$, I. Yang$^{*,+}$, W. Kang$^{*}$, L.N. Pfeiffer$^{\S}$, K.W. Baldwin$^{\S}$, and  K.W. West$^{\S}$}

\affiliation{$*$James Franck Institute and Department of Physics,
 University of Chicago, Chicago, Illinois 60637\\
 $+$Korea Research Institute of Standards and Science
1 Doryoung-Dong, Yuseong-Gu, Daejeon, Rep. of Korea\\
${\S}$Bell Laboratories, Lucent Technologies, 600 
Mountain Avenue, Murray Hill, NJ 07974}

\date{\today}

\begin{abstract}
Tunneling characteristics of a two-dimensional lateral tunnel junction (2DLTJ) are reported. A pseudogap on the order of Coulomb energy is detected in the tunneling density of states (TDOS) when two identical two-dimensional electron systems are laterally separated by a thin energy barrier. The Coulombic pseudogap remains robust well into the quantum Hall regime until it is overshadowed by the cyclotron gap in the TDOS. The pseudogap is modified by in-plane magnetic field, demonstrating a non-trivial effect of in-plane magnetic field on the electron-electron interaction. 

\end{abstract}


\maketitle

Tunneling has been demonstrated to be a valuable tool in the study of collective dynamics of electrons in low dimensions. In response to a tunneling event, electrons in the conduction band must adjust themselves to accommodate excess electrons and holes produced during the process. Electrical responses of tunnel junctions thereby serve to reveal the correlation properties of many-electron systems. In considering the effects of electron-electron interaction in low-dimensional tunnel junctions, Altshuler and Aronov first pointed out that the perturbative effects of interaction lead to a suppression of the tunneling density of states (TDOS) at the Fermi edge compared to that in the non-interacting limit\cite{Altshuler80,Altshuler85}. Tunneling conductance, which provides a measure of the TDOS, consequently exhibits a minimum at zero bias voltage, leading to the phenomenon of zero-bias anomalies in tunnel junctions. To date, a number of metallic and semiconductor tunnel junctions have provided experimental confirmation of the zero-bias anomaly in low dimensions\cite{Dynes81,Imry82,Gershenson85,Hsu94,Massey96,Pierre01}.

In the limit of strong electron-electron correlation, the number of available tunneling states at the Fermi edge becomes sharply curtailed, and a pseudogap emerges below some characteristic Coulomb energy in the TDOS. A well-known example of such a correlation-driven modification of the single particle density of states involves formation of a Coulomb gap in two- and three-dimensional disordered systems\cite{Efros75,Shklovskii}. Efros and Shklovskii proposed that disorder localizes electrons and limits their ability to screen Coulomb interaction. A soft gap is produced at the Fermi edge in the single particle spectrum if the localization length is considerably smaller than the inter-electron distance. The density of states, $D(E)$, within the Coulomb gap is given by $D(E) \propto |E-E_{F}|^{d-1}$, where $d$ is the dimension. As the interplay of disorder and interaction in two dimensions has received much attention recently\cite{Abrahams01}, study of the Coulomb gap and the associated TDOS of a two-dimensional electron system provides a valuable spectroscopy of electron correlation in two dimensions.

In this paper, we report on the low temperature tunneling characteristics of a two-dimensional lateral tunnel junction (2DLTJ) that couples two side-by-side two-dimensional electron systems across a thin rectangular tunnel barrier. Our experiment provides a clear evidence of a pseudogap in the TDOS induced by strong electron-electron correlation in the 2DLTJ. The pseudogap regime, within which the tunneling conductance varies linearly with bias voltage, is distinguished from the ohmic behavior found at larger bias voltages by a set of conductance maxima occurring at a bias energy comparable to the Coulomb energy of the two-dimensional electron system. The pseudogap features remain robust well into the quantum Hall regime in the presence of the magnetic field perpendicular to the two-dimensional electron system. In-plane magnetic fields, on the other hand, induce an unexpected weakening of the pseudogap, pointing to some non-trivial effect on the Coulomb interaction of the two-dimensional electron systems.

\begin{figure}
\includegraphics[width=3.5in]{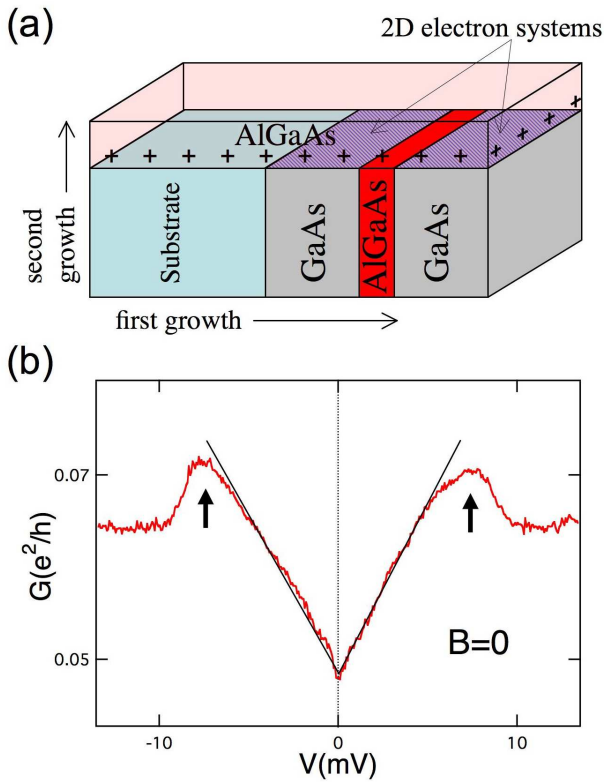}
\caption{\label{fig:dcbfig1} (a) Layout of the two-dimensional lateral tunnel junction grown by cleaved edge overgrowth. (b) Differential conductance as a function of voltage bias at zero magnetic field and a temperature of 300 mK.}
\end{figure}

Fig.~\ref{fig:dcbfig1}a illustrates the layout of the 2DLTJ which incorporates an 8.8-nm-thick Al$_{0.3}$Ga$_{0.7}$As between two coplanar two-dimensional sheets of electrons. Junctions were fabricated through cleaved edge overgrowth\cite{Pfeiffer90,Kang00}. Initial growth along the (100) direction consists of an undoped 13-$\mu$m GaAs followed by an 8.8-nm-thick Al$_{0.3}$Ga$_{0.7}$As, and completed by a 14-$\mu$m layer of undoped GaAs. Then the entire structure is subsequently cleaved along the (110) plane and a modulation doping is performed over the exposed edge. Resulting two side-by-side sheets of identical two-dimensional electron systems are separated from each other by the Al$_{0.3}$Ga$_{0.7}$As barrier. Photolithography is performed to make independent contacts to two-dimensional electron systems on both sides of the barrier. Samples with areal electron densities of $n_{1}\sim1.1\times 10^{11}$ cm$^{-2}$ and $n_{2}\sim7\times 10^{10}$ cm$^{-2}$ were studied. (110) monitor wafers yield typical mobilities of about 5$\times$ 10$^{5}$ cm$^{2}$/Vs. The barrier width and height were chosen to ensure that the transport across the barrier is in the weak tunneling regime.

Fig.~\ref{fig:dcbfig1}b shows the differential conductance $G=dI/dV$ across a 2DLTJ under zero magnetic field and a temperature of 300 millikelvin. The conductance is symmetric about the sharply defined zero-bias minimum and increases linearly with the bias voltage until a set of conductance peaks appears around  $\pm 7.2$ mV. The tunneling becomes ohmic above the $\pm 7.2$ mV peaks, demonstrated by the constant conductance under larger bias voltages. The observed behavior about the zero bias is consistent with the expectation for a pseudogap structure in the TDOS due to the presence of strong electron-electron interaction. The conductance maxima located around $\pm7.2$ mV can be interpreted as resonances in conductance at the edges of the pseudogap,  providing a measure of the magnitude of the pseudogap in the TDOS.  Since the width of the tunnel barrier (8.8 nm) is considerably smaller than the average inter-electron distance $a = 2(\pi n)^{-1/2}$ of 34 nm for the density $n$ = $1.1\times 10^{11}$ cm$^{-2}$, it is expected that Coulomb interaction plays a predominant role in determining the tunneling dynamics of 2DLTJs. 

\begin{figure}
\includegraphics[width=3.5in]{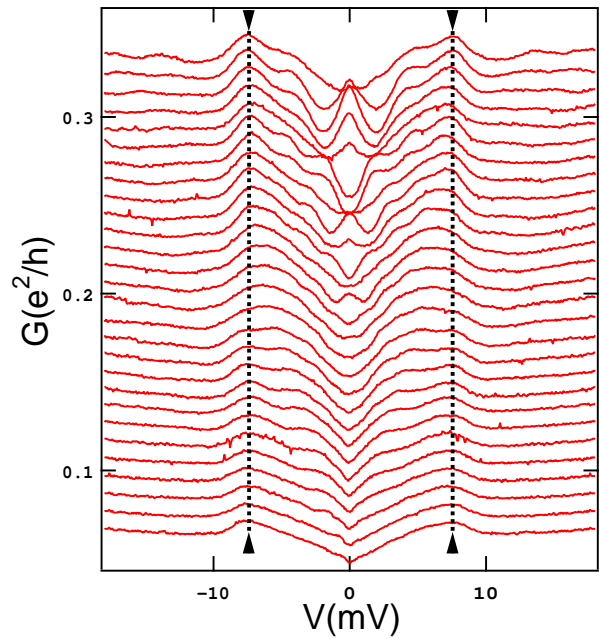}
\caption{\label{fig:dcbfig2} Low temperature tunneling conductance between zero magnetic field and B = 1.35 tesla in steps of 0.05 tesla. The curves have been vertically offset for clarity. Dashed lines indicate the conductance peaks around $\pm$7.2 mV that flank the zero-bias anomalies due to the pseudogap.}
\end{figure}

The significance of the pseudogap structure at $\pm 7.2$ mV is further illustrated by the effect of magnetic field on the 2DLTJ. Fig.~\ref{fig:dcbfig2} shows the evolution of the conductance across the 2DLTJ under perpendicular magnetic fields from 0 to 1.35 tesla in increments of 0.05 tesla. In the range of magnetic field studied, the conductance at zero bias is generally suppressed by 30 $\%$ $\sim$ 70 $\%$ relative to the conductance maxima around $\pm$7.2 mV. This suppression points to the significance of electron-electron interaction in 2DLTJs as similarly reported in the studies of zero-bias anomalies in low-dimensional tunnel junctions\cite{Altshuler80,Altshuler85,Dynes81,Imry82,Gershenson85,Hsu94,Pierre01}. The close proximity of the two two-dimensional electron systems to each other in the 2DLTJ should produce a significant correction to the TDOS. According to our observations, the zero-bias anomaly and the pseudogap features generally remain unperturbed as the quantum Hall regime is approached. For $B$ = 1.35 tesla, the bulk Landau level filling approaches  $\nu \approx 3$ but only a minor variation of the $\pm$7.2 mV feature is detected. Such robustness of the $\pm 7.2$ mV feature, coupled with the strong suppression of tunneling conductance around zero bias, clearly demonstrates the pseudogap physics  of the 2DLTJ.

In addition to the pseudogap features, a set of small conductance oscillations are found 
in Fig.~\ref{fig:dcbfig2}. These oscillations occur as a result of  tunneling between two counterpropagating edge states sharing equal transverse momentum\cite{Ho94,Takagaki00,Mitra01,Kollar02,Nonoyama02,Kim03} and become prominent in the quantum Hall regime\cite{Kang00,Yang}. We find that the tunnel spectrum evolves gradually from the pseudogap physics near B = 0 to the edge-state tunneling in the quantum Hall regime. The crossover between the two  phenomena occurs  around filling factor $\nu = 3$, where the cyclotron energy is comparable to the magnitude of the pseudogap. Beyond $\nu = 3$, the edge-state tunneling becomes very strong and overshadows the pseudogap features.

In the Coulomb gap envisioned by Efros and Shklovskii, the density of states in two dimensions varies linearly with energy inside the Coulomb gap, $D(E) \propto |E-E_{F}|$\cite{Efros75,Shklovskii}. Because of the relatively high mobility of the two-dimensional electrons in our 2DLTJ, the localization length of our sample is expected to be larger than the sample size, and the applicability of the Coulomb gap picture to our system appears unclear. However, with a barrier thinner than the inter-electron distance, the laterally separated two-dimensional electrons interact with each other across the barrier through a strong Coulomb interaction in the 2DLTJ. It follows that the strong electron-electron interaction across a 2DLTJ is responsible for the correlation-induced pseudogap in the TDOS. In addtion, the tunneling conductance in a 2DLTJ arises from a convolution of the density of states of electrons lying on both sides of the barrier. The linear dependence of the conductance on the bias voltage can be interpreted in terms of the density of states for each side as varying as $D(E) \propto |E-E_{F}|^{1/2}$ for energies below the pseudogap\cite{Gap}. 

Emergence of the pseudogap in the 2DLTJ is evident from the presence of highly correlated electronic motion. Prior to a tunneling event, electrons near the barrier distribute themselves to minimize the overall energy.  In order to transport an electron across the junction, work must be performed to extract the electron from its initial low-energy state, which leaves a vacant hole behind. Injection of the extracted electron into the two-dimensional electron system on the opposite side of the barrier requires further work to overcome the ensuing repulsion between electrons. From this argument, it follows that a tunneling event across a 2DLTJ requires an average correlation energy on the order of Coulomb energy $E_{C}$ to overcome the Coulomb energy barrier against tunneling.  The number of accessible tunneling states at energies below the characteristic Coulomb energy $E_{C}$ is thereby small, but increases substantially at energies close to $E_{C}$. This produces a pseudogap below $E_{C}$ in the TDOS.  For a two-dimensional electron system, the Coulomb energy is given by $E_{C} = e^{2}/\epsilon a$, where $\epsilon$ is the dielectric constant, $a = 2(\pi n)^{-1/2}$ is the interparticle distance, and $n$ is the areal density of the two-dimensional electron system. 

\begin{figure}
\includegraphics[width=3.5in]{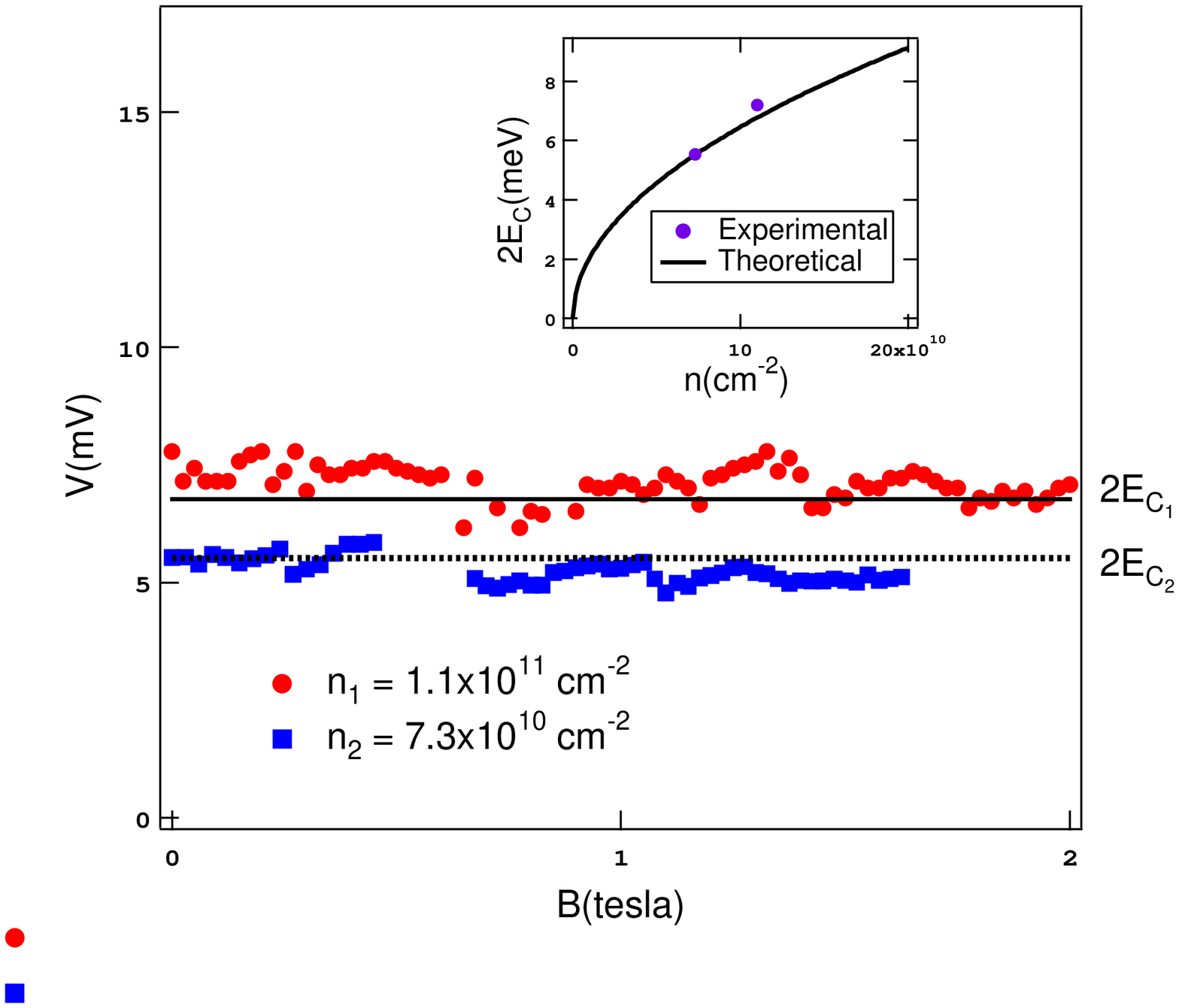}
\caption{\label{fig:dcbfig3} Locations of conductance maxima as a function of perpendicular magnetic field for both $n_{1} = 1.1 \times 10^{11}$ cm$^{-2}$ and $n_{2} = 7.3 \times 10^{10}$ cm$^{-2}$, where $n$'s are the electron densities. Predicted Coulomb energies $2E_{C_{1}}$ = 6.77 meV and $2E_{C_{2}}$ = 5.52 meV are indicated in a solid line and a dashed line respectively. Inset: Calculated $2E_{C}$ as a function of electron density $n$ is shown as a solid curve. Averaged pseudogap magnitudes for the experimental samples are shown as dots.}
\end{figure}

Empirically we find the pseudogap magnitude to be $\sim$2$E_{C}$. Fig.~\ref{fig:dcbfig3} summarizes the positions of the conductance peaks associated with the pseudogaps of two 2DLTJs with different electron densities. The solid and dashed lines respectively correspond to twice the Coulomb energy, $2E_{C}$, for electron densities $n_{1}$ and $n_{2}$. For $n_{1} = 1.1\times 10^{11}$ cm$^{-2}$, the pseudogap energy of $\sim \pm7.2$ meV is comparable to $2E_{C_{1}}$ = 6.77 meV. With the density of $n_{2} = 7.3 \times 10^{10}$ cm$^{-2}$, the pseudogap energy is $\sim \pm$5.4 meV, which is comparable to $2E_{C_{2}}$ = 5.52 meV.  The inset of Fig.~\ref{fig:dcbfig3} shows the square-root dependence of the pseudogap energy on the electron density $n$. The multiplicative factor of 2 in front of $E_{C}$ for the pseudogap energy may be due to the work involved in the extraction as well as the injection of tunneling electrons in the 2DLTJ. It remains to be seen whether a detailed theoretical analysis will yield a pseudogap with a magnitude comparable to our experimental results.

We point out the possibility of multiple tunneling processes coexisting with the pseudogap physics discussed so far. The finite tunneling conductance ($\sim 0.05 e^{2}/h$) at zero bias may be interpreted in terms of parallel tunneling processes in the 2DLTJ. A complete suppression of the TDOS may not occur if midgap tunneling states are present in the range of parameters studied in the present experiment.  Since the 2DLTJ lies in a three-dimensional structure, impurities and disorders within the vertically extended GaAs layers may give rise to 2D-3D or 3D-3D tunneling processes that potentially contribute excess states to the TDOS within the pseudogap at biases in excess of the Fermi energy. In such a case, the $|E-E_{F}|^{1/2}$ dependence of the TDOS for a 2DLTJ discussed earlier may need to be modified. Further experiments should clarify the origin of the midgap tunneling states.

\begin{figure}
\includegraphics[width=3.5in]{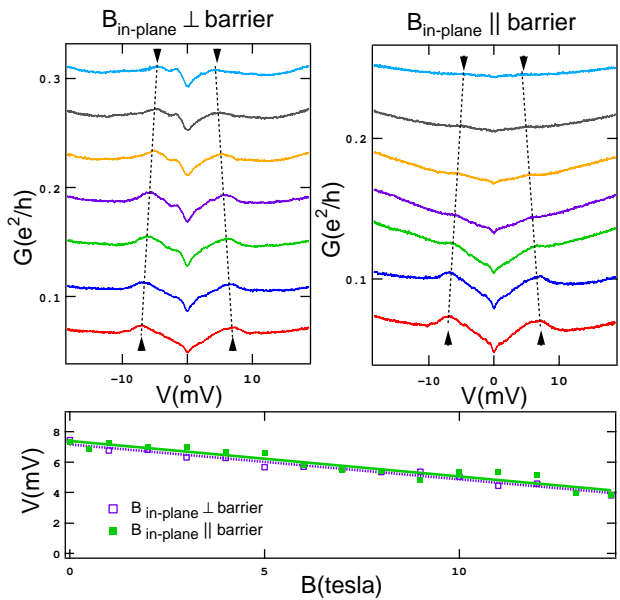}
\caption{\label{fig:dcbfig4} Tunneling characteristics of two-dimensional lateral tunnel junction with in-plane magnetic field (a) perpendicular to the junction and (b) parallel to the junction. The curves have been vertically offset for clarity. (c) Pseudogap magnitude as a function of in-plane magnetic field. Solid and hollow squares are obtained when the in-plane field is respectively perpendicular and parallel to the barrier. Respective linear fits are also shown.}
\end{figure}

We now turn to the effect of in-plane magnetic field on the pseudogap physics of the 2DLTJ. Figs.~\ref{fig:dcbfig4}a and \ref{fig:dcbfig4}b respectively display the evolution of the pseudogap feature as the in-plane magnetic fields perpendicular and parallel to the barrier are increased. When a sufficiently strong in-plane magnetic field is applied, a discernible decrease in the size of the pseudogap can be seen in both cases. In addition, dramatic suppressions of  the zero-bias anomalies and the conductance peaks are observed when the in-plane magnetic field is parallel to the barrier. If the in-plane magnetic field is perpendicular to the  barrier, the gradual suppression of the pseudogap is accompanied by an emergence of tunneling structures at intermediate voltage biases that appears to accentuate the zero-bias anomaly and result in asymmetry in the tunneling characteristics.

This startling difference between the tunneling spectra of the two respective in-plane magnetic field orientations can be understood by considering the effect of Lorentz force on the electrons tunneling across the barrier. With the in-plane magnetic field perpendicular to the barrier, the electrons tunneling across the junction experience no additional force as there is no Lorentz force associated with the magnetic field. When the magnetic field is parallel to the barrier, however, the tunneling electrons experience a Lorentz force along the $z$ direction, perpendicular to the plane. This induced $z$ motion reduces the number of electrons that partake in the tunneling within the plane of two-dimensional electron systems. Resulting reduction in the matrix elements should drastically suppress the tunneling when the magnetic field is parallel to the barrier in comparison to the case with the magnetic field perpendicular to the barrier.

Fig.~\ref{fig:dcbfig4}c summarizes the dependence of pseudogap structure on the magnitude of in-plane magnetic fields parallel and perpendicular to the tunnel barrier. In both orientations the pseudogap decreases at a rate of $\sim$ 0.2 meV/tesla between 0 and 14 tesla.  For the magnitude reduction of the pseudogap we considered the effect of the change in the thickness of the two-dimensional electron systems under in-plane magnetic fields. The extent of the electronic wave functions along the $z$ direction affects the strength of the Coulomb interaction in two-dimensional electron systems\cite{Cooper97}. Application of magnetic fields generally reduces the $z$ extent\cite{Zhang86,Morf02,Jungwirth93}, leading to an enhancement of the Coulomb interaction energy at higher magnetic fields. Since a suppression rather than an amplification of the pseudogap is observed, it appears that in-plane magnetic fields generate some non-trivial effect besides the thickness effect on the electron-electron interaction.  Further theoretical analysis is required to understand the effect of in-plane magnetic fields on the electron-electron correlation in a 2DLTJ.

In summary, our experiment on 2DLTJs provides a clear demonstration of a correlation-induced pseudogap in the TDOS. Strong Coulomb interaction modifies the density of states at the Fermi edge and produces a gap with the magnitude approximately twice the Coulomb energy. The TDOS in the pseudogap regime is susceptible to a significant perturbation from the presence of in-plane magnetic fields. The anomalous reduction in the magnitude of the pseudogap points to some non-trivial alteration of the density of states by in-plane magnetic fields.

We acknowledge J.P. Eisenstein for useful discussions. The work at the University of Chicago is supported by NSF DMR-0203679 and NSF MRSEC Program under DMR-0213745.

\end{document}